%% file: aanda.tex
\documentclass{aa}
\usepackage{graphicx,natbib}
\begin{document}
 
\input macros
\baselineskip=12pt

\title{Improved discretization of the wavelength derivative term
in CMF operator splitting numerical radiative transfer}
\titlerunning{Wavelength Discretization in CMF}
\authorrunning{Hauschildt \& Baron}
\author{Peter H. Hauschildt\inst{1,2} and  E.~Baron\inst{3,4,5}}

\institute{Hamburger Sternwarte, Gojenbergsweg 112, 21029 Hamburg, Germany;
yeti@hs.uni-hamburg.de \and
Dept. of Physics and Astronomy \& Center for
Simulational Physics, University of Georgia, Athens, GA 30602-2451 USA;
\and
Dept. of Physics and Astronomy, University of
Oklahoma, 440 W.  Brooks, Rm 131, Norman, OK 73019 USA;
baron@nhn.ou.edu
\and
NERSC, Lawrence Berkeley National Laboratory, MS
50F-1650, 1 Cyclotron Rd, Berkeley, CA 94720-8139 USA
\and
Laboratoire de Physique Nucl\'eaire et de Haute Energies, CNRS-IN2P3,
University of Paris VII, Paris, France. 
}

   \date{Accepted Dec 17, 2003}

\abstract{We describe two separate wavelength discretization schemes
  that can be used in the numerical solution of the comoving frame
  radiative transfer equation. We present an improved second order
  discretization scheme and show that it leads to significantly less
  numerical diffusion than previous scheme. We also show that due to
  the nature of the second order term in some extreme cases it can
  become numerically unstable. We stabilize the scheme by introducing
  a mixed discretization scheme and present the results from several
  test calculations.
}

\maketitle

\section{Introduction}

The numerical solution of the radiative transfer equation plays a
large role in our interpretation of spectroscopic data of
astrophysical sources. New methods and faster computers have led to a
resurgence of interest in solving the transfer equation 
\citep*[see for example ][]{tuebproc03}. 

The numerical radiative transfer in the co-moving frame (CMF) method
discussed in \cite{s3pap} uses a discretization of the $\partial
\lambda I/\partial \lambda $ terms in the RTE to obtain a formal
solution.  We show that a second order  discretization scheme for the
wavelength derivative leads to better numerical accuracy for a number
of applications, such as stellar winds. In addition, the new
discretization allows us to include the effects of the wavelength
derivative in the construction of the approximate $\Lambda$ operator
used in the operator splitting method.  This improves the
computational performance of the algorithm.  It is possible to mix the
two discretization scheme to tailor the performance of the algorithm
to the problem being considered.  In the following we describe the new
discretization and the construction of the $\lstar$ matrix for
arbitrary bandwidths and  discuss some test results. We have
implemented this scheme into the \texttt{PHOENIX} code 
\citep[see for example][]{jcam}.

\section{Method}

In the following discussion we use the same notation as in
\cite{s3pap} and  reproduce the key equations for convenience.  We
use the spherically symmetric form of the special relativistic, time
independent ($\partial/\partial t \equiv 0$) RTE (hereafter, SSRTE),
the restriction to plane parallel geometry is straightforward. The
calculation of the characteristics is identical that of \citet{s3pap} and we
thus assume that the characteristics are known.  First, we will
describe the process for the formal solution, then we will describe
how we construct the approximate $\L$ operator, $\lstar$.

\subsection{Radiative Transfer Equation}

In the CMF the SSRTE
in the wavelength ($\lambda$) scale is given by \cite[]{FRH,s3pap}
\begin{eqnarray}
   a_r \pder{I}{r} + a_\mu \pder{I}{\mu} 
    + a_\lambda\pder{\lambda I}{\lambda} + 4a_\lambda I = 
   \eta - \chi I         
\label{ssrte}                                     
\end{eqnarray}
with
\begin{eqnarray}
     a_r       &= &\gamma(\mu+\beta) ,              \\
     a_\mu     &= &\gamma(1-\mu^2)
                   \left[\div{1+\beta\mu}{r} 
                          - \gamma^2\left(\mu+\beta\right)\pder{\beta}{r}
                   \right] ,                       \\
     a_\lambda &= &\gamma 
                   \left[ \div{\beta(1-\mu^2)}{r}
                         +\gamma^2\mu\left(\mu+\beta\right)\pder{\beta}{r}
                   \right] .                       
\end{eqnarray}
Along the characteristics, Eq.~\ref{ssrte} has the form \citep{mih80}
\begin{equation}
\frac{dI_l}{ds} + a_l\frac{\partial \l I}{\partial \l} = \eta_l - 
(\chi_l+4a_l)I_l \label{rte-char}
\end{equation}
where $ds$ is a line element along a (curved) characteristic, $I_l(s)$ 
is the specific intensity along the characteristic at point $s\ge 0$ ($s=0$
denotes the beginning of the characteristic) and wavelength point $\l_l$.
The coefficient $a_l$ is defined by
\[
     a_l = \gamma  
                   \left[ \frac{\beta(1-\mu^2)}{r}
                         +\gamma^2\mu\left(\mu+\beta\right)\pder{\beta}{r}
                   \right]                        
\]
where $\beta=v/c$, $\gamma=\sqrt{1-\beta^2}$ and $r$ is the radius.
$\eta_l$ and $\chi_l$ are the emission and extinction coefficients 
at wavelength $\l_l$, respectively.

\subsubsection{First $\partial I/\partial \lambda$ discretization}
As in \citet{FRH} and \citet{s3pap}, we discretize the wavelength derivative
in the SSRTE with a fully implicit method in order to ensure stability:
\[
     \left.\pder{\lambda I}{\lambda}\right|_{\lambda=\lambda_l}
   =
     \div{\lambda_l I_{\lambda_l} - \lambda_{l-1} I_{\lambda_{l-1}}}
         {\lambda_l - \lambda_{l-1}}
\]
so that the wavelength-discretized SSRTE becomes
\[
    \div{dI_{\lambda_l}}{ds} +  a_\lambda 
        \div{\lambda_l I_{\lambda_l} - \lambda_{l-1} I_{\lambda_{l-1}}}
            {\lambda_l - \lambda_{l-1}}
    = \eta_{\lambda_l} - (\chi_{\lambda_l}+4a_\lambda)I_{\lambda_l}.
\]

If we now define the optical depth scale along the ray as
\[
     d\tau \equiv \chi
        +a_\lambda \left(4+\div{\lambda_l}{\lambda_l-\lambda_{l-1}}\right)
     \equiv \hat\chi ds,
\]
and introduce the source function, $S=\eta/\chi$, we have
\[
   \div{d I}{d\tau} 
 = 
   I - \div{\chi}{\hat\chi} 
       \left(S + \div{a_\lambda}{\chi}\div{\lambda_{l-1}}{\lambda_l-\lambda_{l-1}}
                 I_{\lambda_{l-1}}\right) 
 \equiv I-\hat S .
\]

With this definition, 
the formal solution of the SSRTE along the characteristics can be written
in the following way \citep[cf.][for a derivation of the formulae]{ok}:
\begin{eqnarray}
       I(\tau_i) &=& I(\tau_{i-1}) \exp(\tau_{i-1}-\tau_i)\\\nonumber
                 &&  +\int_{\tau_{i-1}}^{\tau_i} \hat S(\tau) \exp(\tau_{i-1}-\tau)
                      \, d\tau \\ 
       I(\tau_i) &\equiv& I_{i-1}\exp(-\Delta\tau_{i-1})+\Delta I_i   
\end{eqnarray}
where we have suppressed the index labeling the ray; $\tau_i$ denotes the
optical depth along the ray with $\tau_1\equiv 0$ and $\tau_{i-1} \le \tau_i$
while $\tau$ is calculated using piecewise linear interpolation of 
$\hat\chi$ along the ray, viz.\
\begin{eqnarray}
     \Delta\tau_{i-1} = (\hat\chi_{i-1}+\hat\chi_i)|s_{i-1}-s_i|/2. \label{taudef}
\end{eqnarray}
The source function $\hat S(\tau)$ along a characteristic is interpolated by 
linear or parabolic polynomials so that 
\begin{eqnarray}
    \Delta I_i = \alpha_i \hat S_{i-1} 
                + \beta_i \hat S_i + \gamma_i \hat S_{i+1},
\end{eqnarray}
$\Delta\tau_i \equiv \tau_{i+1}-\tau_i$ is the optical depth along the
ray from point $i$ to point $i+1$. The coefficients $\alpha_i$, $\beta_i$,
and $\gamma_i$ are given in \citet{ok}.


For the tangent rays \citep[for example ray 1 in Fig.~1 of][]{s3pap},
the formal solution starts at point
2 with $I_1$ given as the outer boundary condition and proceeds 
along the ray.
The formal solution for a core-intersecting ray is split into
two parts:
(i) integration from point 1 to point $N$, where $I_1$ is given as
the outer boundary condition and
(ii) integration from point $N+2$ to point $2N$, where $I_{N+1}$ is
given as the inner boundary condition.

\begin{figure}
\centering
\includegraphics[height=0.4\vsize]{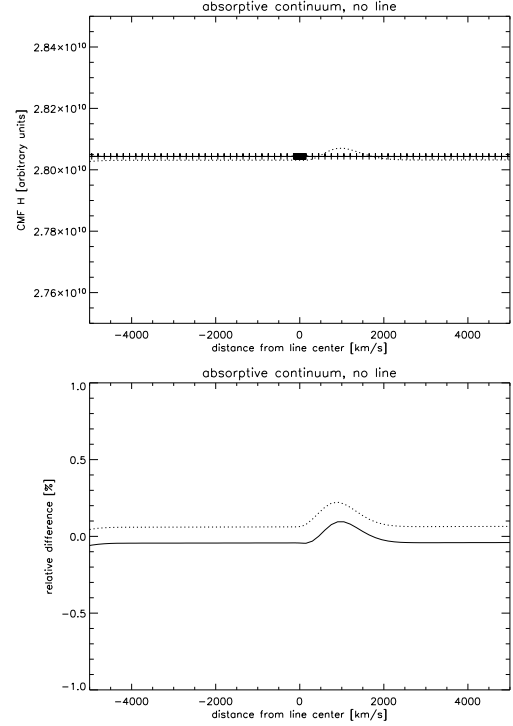}
\caption[]{\label{abs.cont}Absorptive continuum ($\epsilon_c=1$). Top
part: CMF H in arbitrary units (full line: $\zeta = 0$, 
dotted line: $\zeta = 1$), bottom part: relative differences between
CMF H (full line) and CMF J (dotted line) for the two discretizations. The $+$ 
signs give the location of the actual wavelength points used in the computation.}
\end{figure}

\subsubsection{Second $\partial I/\partial \lambda$ discretization}

The discretization of the $\partial I/\partial \lambda$ derivative in
the previous section can be deferred.
We first rewrite Eq.~\ref{rte-char} as


\def\plm{p_{l,l-1}}
\def\plp{p_{l,l+1}}
\def\pll{p_{l,l}}
\def\Ilm{I_{l-1}}
\def\Ilp{I_{l+1}}
\def\Ill{I_{l}}

\begin{eqnarray}
\frac{dI_l}{d\tau} - \frac{a_l}{\hat\chi_l}\frac{\partial \l I}{\partial \l} =
I-\frac{\eta_l}{\hat\chi_l} \label{rte-tau}
\end{eqnarray}
with 
\[
\hat\chi_l = \chi_l+4a_l
\]
and the definition of the CMF optical depth
\[
d\tau \equiv -\hat\chi_l ds
\]
along the characteristic.
To obtain an expression for the formal solution, we rewrite Eq.~\ref{rte-tau}
as
\[
\frac{dI}{d\tau} = I - \hat S - \tilde S 
\]
with
\begin{eqnarray}
\hat S &= &\frac{\chi}{\hat\chi_l}S = \frac{\eta_l}{\hat\chi_l} \\
\tilde S &= &-\frac{a_l}{\hat\chi}\frac{\partial \l I}{\partial \l} \\
\end{eqnarray}


\def\Imm{I_{i-1,l-1}}
\def\Imp{I_{i-1,l+1}}
\def\Ipm{I_{i+1,l-1}}
\def\Ipp{I_{i+1,l+1}}
\def\Iil{I_{i,l}}
\def\Iml{I_{i-1,l}}

\def\mm{_{i-1,l-1}}
\def\mp{_{i-1,l+1}}
\def\pm{_{i+1,l-1}}
\def\pp{_{i+1,l+1}}
\def\il{_{i,l}}
\def\im{_{i,l-1}}
\def\ip{_{i,l+1}}
\def\ml{_{i-1,l}}
\def\pl{_{i+1,l}}

With this we obtain the following expression for
the formal solution \cite[see also Eq.~20 in][]{s3pap}
\begin{equation}
\Iil = \Iml\exp(-\Delta\tau_{i-1}) + \delta\hat\Iil + \delta\tilde\Iil
\label{fs}
\end{equation}
with the definitions
\[
\delta\hat\Iil = \alpha\il \hat S\ml + \beta\il \hat S\il + \gamma\il \hat S\pl
\]
and
\[
\delta\tilde\Iil = \tilde\alpha\il \tilde S\ml + \tilde\beta\il \tilde S\il 
\]
The index $i$ labels the (spatial) points along a characteristic, the index
$l$ denotes the wavelength point. The coefficients $\alpha\il$, $\beta\il$,
and $\gamma\il$ are given in \citet{s3pap} and \citet{ok}, here they are 
calculated for a fixed wavelengths for all points along a characteristic.
$\hat S$ is a vector of known quantities (the old mean intensities and 
thermal sources). The $\tilde S$ contain the effects of the velocity field
on the formal solution and are given by
\begin{eqnarray*}
\tilde S\ml & = & -\frac{a\ml}{\hat\chi\ml} 
\left.\frac{\partial \l I}{\partial \l}\right|_{\ml}\\
\tilde S\il & = & -\frac{a\il}{\hat\chi\il} 
\left.\frac{\partial \l I}{\partial \l}\right|_{\il}
\end{eqnarray*}
Note that the integration of $\tilde S$ is performed using linear elements 
in order to allow for a recursive formal solution. Parabolic elements can be
used but require the solution of matrix equations to obtain the formal
solution. 

If the velocity field is monotonically increasing
or decreasing we use a stable upwind 
discretization of the wavelength derivative. In both cases, the problem
becomes an initial value problem and can be solved for each wavelength once
the results of the previous (smaller or longer) wavelength points are known.
For a monotonically increasing velocity field this gives
\begin{eqnarray}
\tilde S\ml & = & -\frac{a\ml}{\hat\chi\ml} 
\left[ \frac{\l_{l}}{\l_l-\l_{l-1}} I\ml\right.\\
&&-\left.\frac{\l_{l-1}}{\l_l-\l_{l-1}} I\mm \right]\\
\tilde S\il & = & -\frac{a\il}{\hat\chi\il} 
\left[ \frac{\l_{l}}{\l_l-\l_{l-1}} I\il\right.\\
&& -\left.\frac{\l_{l-1}}{\l_l-\l_{l-1}} I\im \right]
\end{eqnarray}
Here, and in the following, we use a sorted wavelength grid with
$\l_{l-1}<\l_l<\l_{l+1}$ for monotonically increasing velocities (the order
of the grid would be reversed for monotonically decreasing velocity fields).
For non-monotonic velocity fields this is no longer the case and the formal
solution needs to explicitly account for the wavelength couplings
\citep{bh03}.

With these formulae we can write $\delta\tilde\Iil$ in the form
\begin{eqnarray*}
\delta\tilde\Iil &=&
\tilde\alpha\il\left(p\ml I\ml - p\mm I\mm \right)\\
&& +
\tilde\beta\il \left(p\il I\il - p\im I\im \right)
\end{eqnarray*}
where
\begin{eqnarray*}
p\ml & = & -\frac{\l_{l}}{\l_l-\l_{l-1}}\frac{a\ml}{\hat\chi\ml} \\
p\mm & = & -\frac{\l_{l-1}}{\l_l-\l_{l-1}} \frac{a\ml}{\hat\chi\ml}\\
p\il & = & -\frac{\l_{l}}{\l_l-\l_{l-1}}\frac{a\il}{\hat\chi\il}\\
p\im & = & -\frac{\l_{l-1}}{\l_l-\l_{l-1}}\frac{a\il}{\hat\chi\il}\\
\end{eqnarray*}
The formal solution then assumes the form
\begin{eqnarray*}
\left(1-\tilde\beta\il p\il\right)I\il &=& 
\left(\tilde\alpha\il+\exp\left(-\Delta\tau_{i-1}\right)\right)I\ml\\
&&-\tilde\alpha\il p\mm I\mm \\
&&- \tilde\beta\il  p\im I\im
+\delta\hat\Iil
\end{eqnarray*}
This equation can be solved recursively along a characteristic
to calculate the $I\il$ for all $i$ and fixed $l$. The form of the
wavelength discretization is now second order accurate in $\Delta\lambda$.

The new form of the formal solution requires only small changes
to the construction of the $\lstar$ operator discussed in
\citet{s3pap} and \citet*{aliperf}.

We describe the construction of $\lstar$ for arbitrary bandwidth using
the example of a tangential characteristic.  The intersection points
(including the point of tangency) are labeled from left to right, the
direction in which the formal solution proceeds.  For convenience, we
label the characteristic tangent to shell $k+1$ as $k$. Therefore, the
characteristic $k$ has $2k+1$ points of intersection with discrete
shells $1\ldots k+1$.  To compute row $j$ of the discrete
$\Lambda$-operator (or $\Lambda$-matrix), $\Lambda_{ij}$, we
sequentially label the intersection points of the characteristic $k$
with the shell $i$ and define auxiliary quantities $\lambda_{ij}^k$
and $\hat\lambda_{ij}^k$ as follows:
\[
\begin{array}{lcl}
\multicolumn{3}{c}{{\rm\ for\ } i<j-1}\\
\lambda^k_{i,j} & = & 0  \\
\multicolumn{3}{c}{{\rm\ for\ } i=j-1} \\
\lambda^k_{j-1,j} & = & \left(1-\tilde\beta^k_{j-1} p^k_{j-1,l}\right)^{-1}\gamma^k_{j-1}\\
\multicolumn{3}{c}{{\rm\ for\ } i=j} \\
\lambda^k_{j,j} & = & \left(1-\tilde\beta^k_{j} p^k_{j,l}\right)^{-1}\\
&&\left( \lambda^k_{j-1,j}\left[\tilde\alpha^k_{j}
  p^k_{j-1,l}+\exp(-\Delta\tau^k_{j-1})\right]
 + \beta^k_{j} \right)\\
\multicolumn{3}{c}{{\rm\ for\ } i=j+1} \\                                      
\lambda^k_{j+1,j} & = & \left(1-\tilde\beta^k_{j+1}
p^k_{j+1,l}\right)^{-1}\\
&& \left( \lambda^k_{j,j}\left[\tilde\alpha^k_{j+1}
  p^k_{j,l}+\exp(-\Delta\tau^k_{j})\right]  + \alpha^k_{j+1}
\right)\\ 
\multicolumn{3}{c}{{\rm\ for\ } j+1 < i \le k+1} \\                                      
\lambda^k_{i,j} & = & \left(1-\tilde\beta^k_{i}
p^k_{i,l}\right)^{-1}\\
&& \left( \lambda^k_{i-1,j}\left[\tilde\alpha^k_{i}
  p^k_{i-1,l}+\exp(-\Delta\tau^k_{i-1})\right] \right) 
\end{array}
\]
For the calculation of $\hat\lambda^k_{i,j}$, we obtain:
\[
\begin{array}{lcl}
\multicolumn{3}{c}{{\rm\ for\ } i = k+2}  \\
\hat\lambda^k_{i,j} & = & \left(1-\tilde\beta^k_{i}
p^k_{i,l}\right)^{-1}\\
&&\left(\lambda^k_{i-1,j}\left[\tilde\alpha^k_{i}
  p^k_{i-1,l}+\exp(-\Delta\tau^k_{i-1})\right] \right) \\
\multicolumn{3}{c}{{\rm\ for\ } k+2 <i < k+j+2}  \\
\hat\lambda^k_{i,j} & = & \left(1-\tilde\beta^k_{i}
p^k_{i,l}\right)^{-1}\\
&& \left(\hat\lambda^k_{i-1,j}\left[\tilde\alpha^k_{i}
  p^k_{i-1,l}+\exp(-\Delta\tau^k_{i-1})\right] \right) \\
\multicolumn{3}{c}{{\rm\ for\ } i = k+j+2}  \\
\hat\lambda^k_{i,j} & = & \left(1-\tilde\beta^k_{i}
p^k_{i,l}\right)^{-1}\\
&& \left(\hat\lambda^k_{i-1,j}\left[\tilde\alpha^k_{i}
  p^k_{i-1,l}+\exp(-\Delta\tau^k_{i-1}) \right] 
                          + \alpha^k_i \right)\\
\multicolumn{3}{c}{{\rm\ for\ } i = k+j+3}  \\
\hat\lambda^k_{i,j} & = & \left(1-\tilde\beta^k_{i}
p^k_{i,l}\right)^{-1}\\
&& \left(\hat\lambda^k_{i-1,j}\left[\tilde\alpha^k_{i} p^k_{i-1,l}+
   \exp(-\Delta\tau^k_{i-1}) \right] 
                          + \beta_i \right)\\
\multicolumn{3}{c}{{\rm\ for\ } i = k+j+4}  \\
\hat\lambda^k_{i,j} & = & \left(1-\tilde\beta^k_{i}
p^k_{i,l}\right)^{-1}\\
&&\left(\hat\lambda^k_{i-1,j}\left[\tilde\alpha^k_{i}
  p^k_{i-1,l}+\exp(-\Delta\tau^k_{i-1}) \right] 
                          + \gamma^k_i \right)\\
\multicolumn{3}{c}{{\rm\ for\ } k+j+5 \le i \le 2k+1}  \\
\hat\lambda^k_{i,j} & = & \left(1-\tilde\beta^k_{i}
p^k_{i,l}\right)^{-1}\\
&&\left(\hat\lambda^k_{i-1,j}\left[\tilde\alpha^k_{i}
  p^k_{i-1,l}+\exp(-\Delta\tau^k_{i-1})\right] \right) 
\end{array}
\]

\noindent Using the $\lambda^k_{ij}$ and $\hat\lambda^k_{ij}$, we can now write the
$\L$-Matrix as
\[
    \Lambda_{ij} = \sum_k \left( \sum_{\{l\}} w^k_{l,j}\lambda^k_{l,j}
                               + \sum_{\{l'\}}
w^k_{2(k+1)-l',j}\hat\lambda^k_{2(k+1)-l',j} 
                           \right)
\]
where $w^k_{i,j}$ are the angular quadrature weights, $\{l\}$ is the set $\{i
\le k+1\}$ and $\{l'\}$ is the set $\{i > k+1\}$.
This expression gives the {\em full} $\L$-matrix, it can easily be 
specialized to compute only certain bands of the $\L$-matrix. In that  case, not all 
of the $\lambda^k_{i,j}$ and  $\hat\lambda^k_{i,j}$ have to be computed, 
reducing the CPU time from that required for
the computation of the full $\L$-matrix. 

\subsubsection{Stability}

In the second discretization scheme, the wavelength derivative
contains an explicit term, which is required  to derive a
recursive method with second order accuracy. We can stabilize the
discretion via a method similar to the standard Crank-Nicholson scheme
\citep{AS72}. 
To combine the two discretization schemes described above we introduce a factor
$\zeta \in [0,1]$ so that for the first scheme in \citet{s3pap} we 
have $\zeta \equiv 1$ and for the second discretization we have $\zeta \equiv 0$.
With this the optical depth scale along the ray $d\tau$ and $\hat\chi$ become
\[
     d\tau = \chi
        +a_\lambda \left(4+\zeta\div{\lambda_l}{\lambda_l-\lambda_{l-1}}\right)
     \equiv \hat\chi ds,
\]
and $\hat S$ is given by
\[
   \hat S
 = 
   \div{\chi}{\hat\chi} 
       \left(S + \zeta\div{a_\lambda}{\chi}\div{\lambda_{l-1}}{\lambda_l-\lambda_{l-1}}
                 I_{\lambda_{l-1}}\right) .
\]
whereas $\tilde S$ becomes
\[
\tilde S = -(1-\zeta)\frac{a_l}{\hat\chi}\frac{\partial \l I}{\partial \l}
\]
With this we can arbitrarily combine the two $\partial I/\partial\lambda$
discretizations by varying $\zeta$ over the interval $[0,1]$. This allows
us to combine them to optimize accuracy of the overall solution of 
the SSRTE.

\subsubsection{Discussion}

It should be expected that the two discretization schemes introduced above
will show different numerical behavior. The $\zeta=1$ scheme handles the 
$I_{\lambda_l}$ and $I_{\lambda_{l-1}}$ parts differently since the (known)
$I_{\lambda_{l-1}}$ is treated like a source term whereas the $I_{\lambda_l}$
term leads to a changed definition of $d\tau$ in the CMF along the ray.
Therefore, the $I_{\lambda_l}$  is basically assumed to vary exponentially
across a characteristic. This will introduce more diffusive behavior 
for large optical depths and strong scattering (where the intrinsic assumptions
for $I_{\lambda_l}$ and $I_{\lambda_{l-1}}$ will differ most). 

This is not the case for the $\zeta = 0$ discretization which assumes
that the $\lambda\partial I/\partial \lambda$ itself can be fitted to
a first order polynomial and treated as a source term. The
discretization step is delayed until the integration of the source
term. Although the discretization is implicit by design, the behavior
of $\partial I / \partial \lambda$ is a priori unknown and the second
order nature of the discretization leads to an explicit term. Under
extreme conditions, e.g., a strong line in an optically thin rapidly
expanding gas, it is possible that the numerical integration is
dominated by the ``known'' term due to large changes in $\partial I /
\partial \lambda$.  This leads to numerical instability.  We therefore
expect that the two discretizations will show different behavior in
numerical tests.

\section{Tests}

In order to compare and test the two different discretization schemes,
we have devised a simple test model. We consider the case of a 2-level
atom with background continuum on a radial grid (assuming spherical
symmetry). The background continuum is assumed to be grey with
adjustable thermalization fraction $\epsilon_c = \kappa_c/\chi_c$ with
$\chi_c=\kappa_c+\sigma_c$.  The line of the 2-level atom is
parameterized by its strength relative to the continuum
$\chi_l/\chi_c$ and its thermalization fraction $\epsilon_l =
\kappa_l/\chi_l$. For the intrinsic line profile we use a Gaussian
profile with a $\Delta \lambda_D$ of $30\kms$.  For the radial
structure, we set $\chi_c \propto r^{-2}$ and use a radial grid with
$R_{\rm out}=101$ and $R_{\rm in} = 1$ (so that the extension of the
test atmosphere is $100$) and a logarithmic optical depth grid from
$\tau = 10^4$ to $10^{-6}$ in the continuum to fix the scale of the
opacities. The velocity law is linear (homologous expansion) with a
prescribed maximum velocity of $1000\kms$. The computations are
performed on a wavelength grid that uses a stepsize of 0.1 line width
inside the line and 5 times the line width outside the line if not
specified in detail for some of the test calculations.

\subsection{Continuum tests}
\begin{figure}
\centering
\includegraphics[height=0.4\vsize]{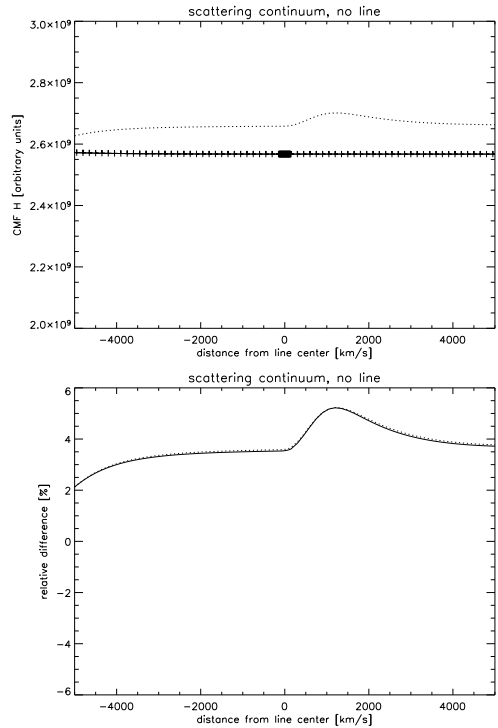}
\caption[]{\label{scatt.cont}Scattering dominated continuum
  ($\epsilon_c=10^{-2}$). Top part: CMF H in arbitrary units (full
  line: $\zeta = 0$, dotted line: $\zeta = 1$), bottom part: relative
  differences between CMF H (full line) and CMF J (dotted line) for
  the two discretizations. The $+$ signs give the location of the
  actual wavelength points used in the computation.}
\end{figure}

\begin{figure}
\centering
\includegraphics[height=0.8\vsize]{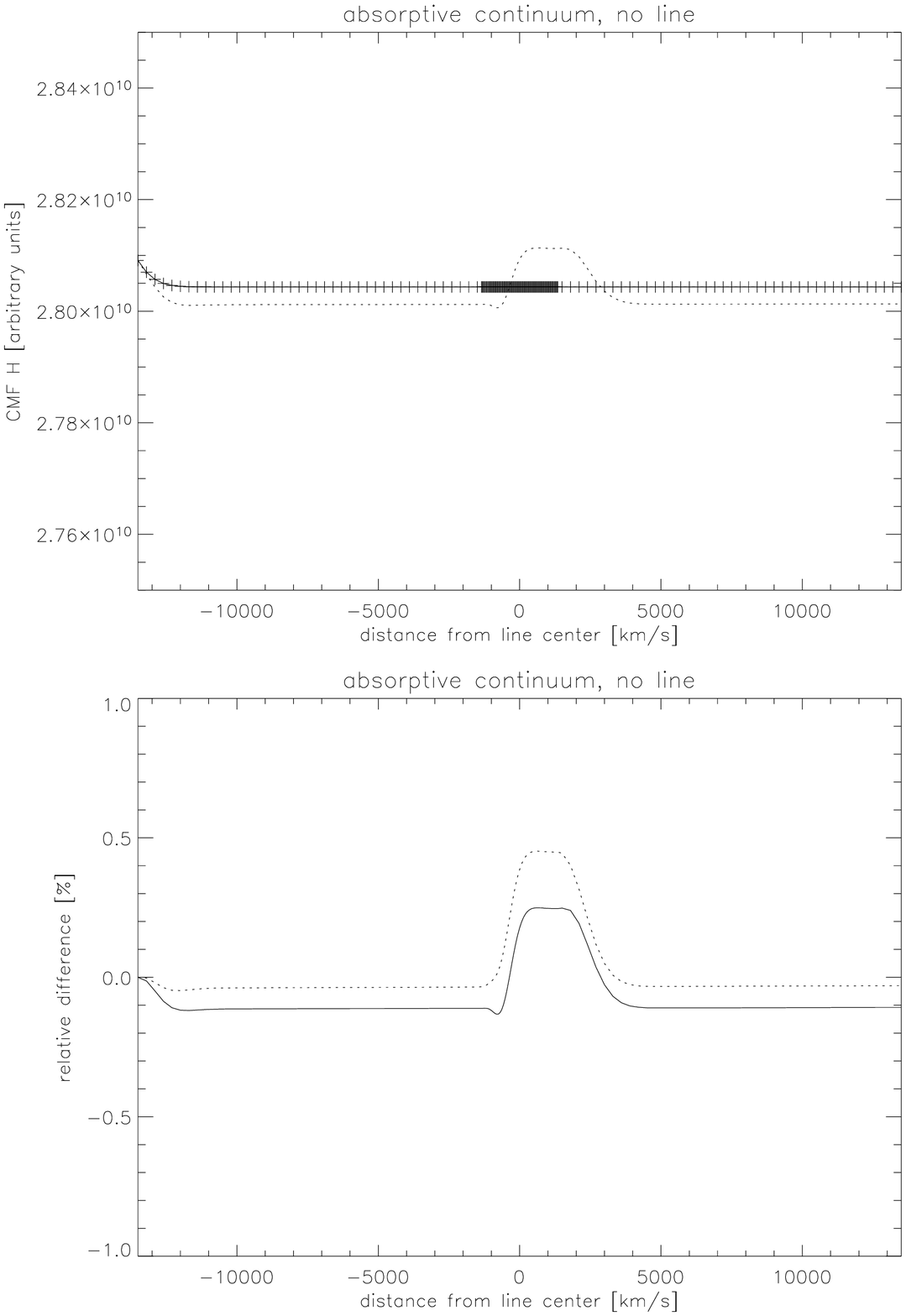}
\caption[]{\label{abs.cont_long}Absorptive continuum
  ($\epsilon_c=1$). Top part: CMF H in arbitrary units (full line:
  $\zeta = 0$, dotted line: $\zeta = 1$), bottom part: relative
  differences between CMF H (full line) and CMF J (dotted line) for
  the two discretizations. The $+$ signs give the location of the
  actual wavelength points used in the computation.}
\end{figure}

\begin{figure}
\centering 
\includegraphics[height=0.8\vsize]{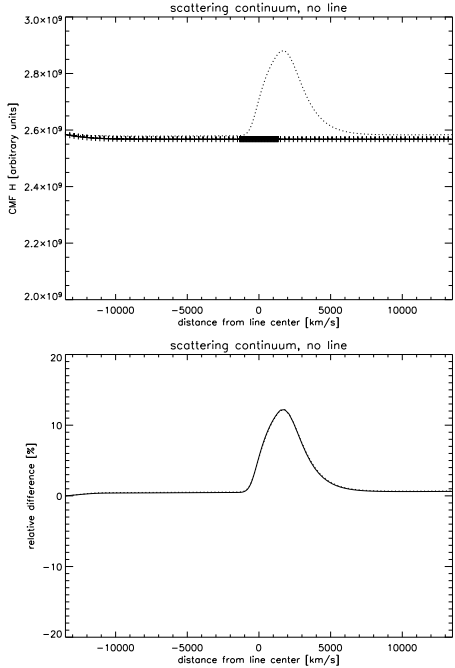}
\caption[]{\label{scatt.cont_long}Scattering dominated continuum
  ($\epsilon_c=10^{-2}$). Top part: CMF H in arbitrary units (full
  line: $\zeta = 0$, dotted line: $\zeta = 1$), bottom part: relative
  differences between CMF H (full line) and CMF J (dotted line) for
  the two discretizations. The $+$ signs give the location of the
  actual wavelength points used in the computation.}
\end{figure}

As a first test we verified that there are no differences between the
solutions for $\zeta = 0$ and $\zeta = 1$ for zero expansion velocity.
The next test is for a pure continuum case (i.e., line strength of
zero).  This is actually a difficult test for the method as the flat
continuum should be reproduced by the algorithm.  The results for
purely absorptive ($\epsilon_c=1$) and scattering dominated
($\epsilon_c=10^{-2}$) continua are shown in Figs.\ref{abs.cont} and
\ref{scatt.cont}, respectively. In this case the CMF $H$ (the first
Eddington moment of the intensity in the CMF) should be constant. From
Fig.~\ref{abs.cont} we see that both discretization schemes deliver
nearly identical results with differences below 0.3\% in the case of
an absorption dominated continuum. For the case of a scattering
dominated continuum the differences are significantly larger, nearly
6\%, cf.~Figure~\ref{scatt.cont}. From the top part of the figure it
is obvious that the differences start to increase just around the
region of resolution change in wavelength at the (zero strength)
line. That scattering increases the diffusive effect of the $\zeta=1$
wavelength discretization as shown by comparing Figs.~\ref{abs.cont}
and \ref{scatt.cont}.  This effect becomes larger and more pronounced
if the wavelength resolution is lower, which is shown in
Figs.~\ref{abs.cont_long} and \ref{scatt.cont_long}.  In this test, we
have decreased the wavelength resolution by roughly a factor of
10. The case of the absorption dominated continuum shows  similar
behavior as the test with higher wavelength resolution with the 
relative differences increasing slightly.  However; the scattering dominated
case 
produces much larger relative differences of up to 13\% between the two
discretizations. Whereas the $\zeta=0$ discretization produces the
expected flat continuum (once the initial conditions in wavelength are
``forgotten''), the $\zeta=1$ discretization produces a pseudo line
feature just due to the change in resolution. This illustrates the
higher accuracy of the second order scheme, which since it is properly
centered produces more accurate results on an irregular grid.

\begin{figure}
\centering
\includegraphics[height=0.8\vsize]{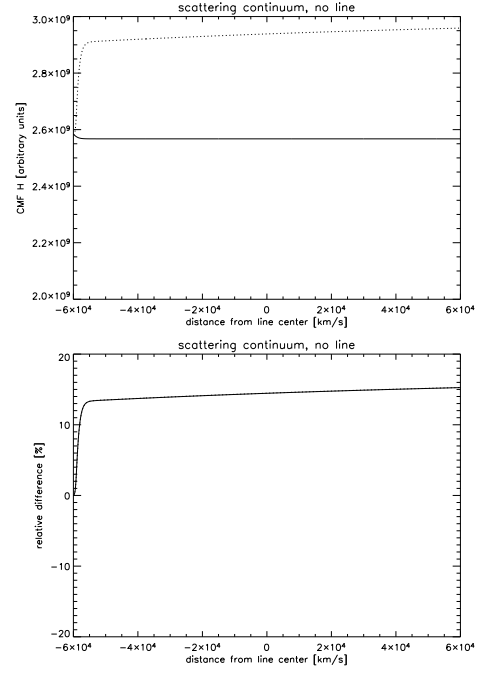}
\caption[]{\label{scatt.cont_wgrid}Scattering dominated continuum
  ($\epsilon_c=10^{-2}$). Top part: CMF H in arbitrary units (full
  line: $\zeta = 0$, dotted line: $\zeta = 1$), bottom part: relative
  differences between CMF H (full line) and CMF J (dotted line) for
  the two discretizations. In this calculation equidistant wavelength
  grid with a spacing of $300\kms$ was used.}
\end{figure}

The case of a constant step size wavelength grid is shown in
Fig.~\ref{scatt.cont_wgrid}.  The step size in wavelength is
$300\kms$. The $\zeta=0$ discretization produces a perfectly flat
continuum after the effect of the (grey) initial condition dies away.
In contrast, the $\zeta=1$ discretization produces considerable
numerical diffusion in this test, on the average the solution deviates
by more than 10\% from the $\zeta=0$ solution. In addition, the continuum
is \emph{not} flat but shows a noticeable increase in CMF $H$ (and $J$)
toward the red.

It is clear that the $\zeta=0$ discretization gives significantly
better results than the $\zeta=1$ approach for the continuum tests.

\subsection{Line tests}

\begin{figure}
\centering
\includegraphics[height=0.8\vsize]{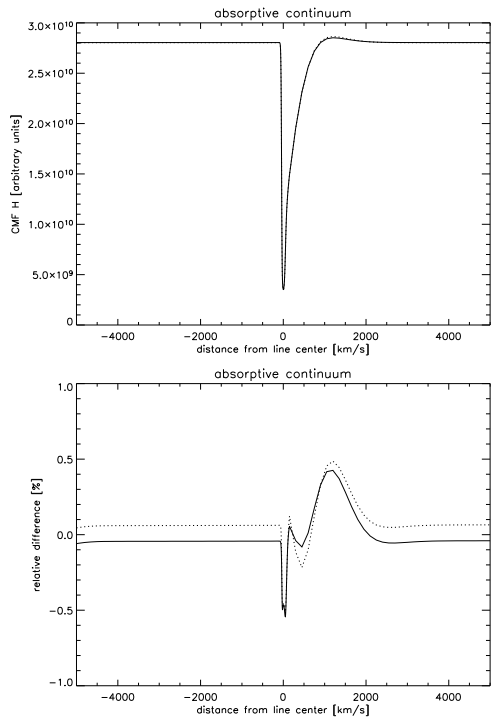}
\caption[]{\label{line.abs.cont}Strong $(\chi_l/\chi_c=100)$
  scattering dominated $(\epsilon_l=10^{-4})$ line with an absorptive
  background continuum ($\epsilon_c=1$). Top part: CMF H in arbitrary
  units (full line: $\zeta = 0$, dotted line: $\zeta = 1$), bottom
  part: relative differences between CMF H (full line) and CMF J
  (dotted line) for the two discretizations.}
\end{figure}

\begin{figure}
\centering 
\includegraphics[height=0.8\vsize]{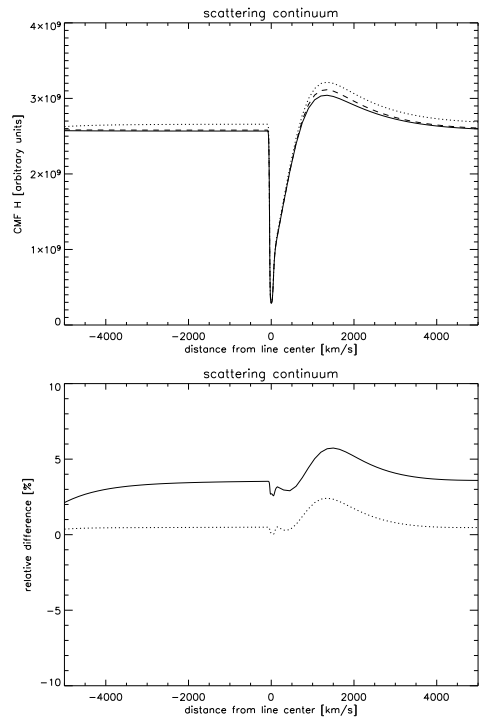}
\caption[]{\label{line.scatt.cont}Strong $(\chi_l/\chi_c=100)$
  scattering dominated $(\epsilon_l=10^{-4})$ line with a scattering
  dominated continuum ($\epsilon_c=10^{-2}$). Top part: CMF H in
  arbitrary units (full line: $\zeta = 0$, dotted line: $\zeta = 1$,
  dashed line: $\zeta = 0.5$), bottom part: relative differences
  between CMF H for the two discretizations with $\zeta = 0$ and
  $\zeta = 1$. In addition, the dotted line shows the relative
  differences between calculations with $\zeta = 0.5$ and $\zeta =
  0$.}
\end{figure}

In contrast to the previous set of tests, we now introduce a strong
line.  We set the line strength $\chi_l/\chi_c=100$ with
$\epsilon_l=10^{-4}$ to simulate a strong, scattering dominated line
of a 2-level atom against a background continuum.
Figs.~\ref{line.abs.cont} and \ref{line.scatt.cont} show the results
for absorption and scattering dominated background continua. In both
cases the relative differences in the continua are comparable to the
previous tests. The differences in the lines are actually
comparatively small in both cases, significantly smaller than might be
expected from the continuum test results considering the fact that the
line is strongly scattering dominated.  This is most likely caused by
the smoothing effect of the rapidly varying opacity across the line.
The effect of continuum scattering is to considerably widen the
line due to line photons being scattered by the continuum.  The more
diffusive $\zeta=1$ discretization scheme produces an emission feature
that is about 2--3\% stronger than the $\zeta=0$ discretization
relative to the continuum. It is interesting to note the strong effect
of even a weak (compared to the line itself) background continuum on
the shape of the line and the differences between the two
discretizations.

For the final test we show in Fig.~\ref{line.scatt.cont} he results of
a calculation for $\zeta = 0.5$. Compared to $\zeta=1$ this
calculation shows much smaller differences from the $\zeta =0$ case in
the continuum.  From the top portion of the figure it is clear that in
the line itself the $\zeta=0.5$ model falls roughly half way in
between the $\zeta=1$ and $\zeta=0$ cases.

In all cases the $\zeta=0$ discretization gives numerically better
results.  However, there are situations where the $\zeta=0$
discretization is numerically unstable, which we discuss in the
following section.

\subsection{Numerical instability for $\zeta=0$}

\begin{figure}
\centering
\includegraphics[angle=90,width=\hsize]{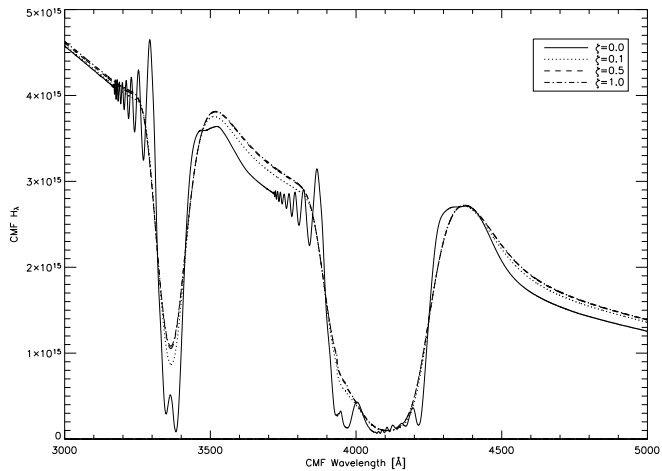}
\caption[]{\label{wiggles_cmf}Example to illustrate the possibility of
numerical instabilities in the $\zeta=0$ discretization. The structure
is taken from an Type Ia supernova calculation, only Ca~II lines are
included in the opacity  and the lines are assumed to be in complete LTE.
}
\end{figure}

\begin{figure}
\centering
\includegraphics[angle=90,width=\hsize]{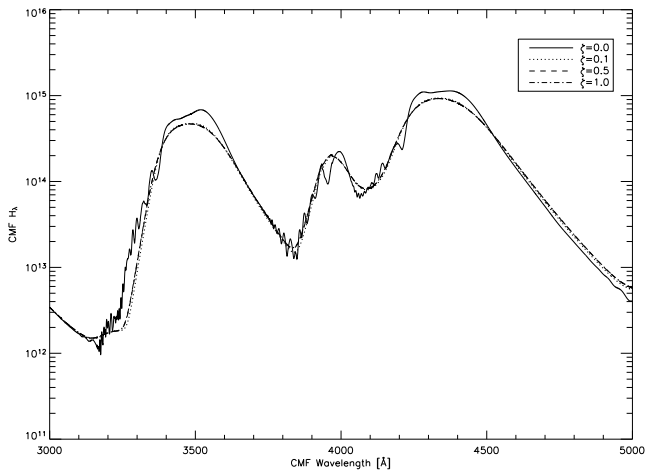}
\caption[]{\label{wiggles_cmf_neb}The same structure as was used in
  Fig.~\ref{wiggles_cmf}, but with nebular (symmetry) boundary
  conditions.
}
\end{figure}

In certain conditions, it is possible for the $\zeta=0$ discretization
to become numerically unstable. This is illustrated in the test case
shown in Fig.~\ref{wiggles_cmf}. The two test lines are in complete
LTE ($S=B$), the background continuum is scattering dominated. The
structure is taken from typical SN Ia structure, but we have
completely ignored the strong line blanketing inherent in SNe Ia by
including line opacity from only Ca~II. This procedure is actually
quite useful for line identification in detailed model
calculations. The velocity field is homologous ($v\propto r$) with a
maximum speed of $30,000\kms$, the inner boundary is at
$v=2,000\kms$. We use this model rather than the simpler more
  academic test cases for two reasons: 1) this is the model where we
  actually discovered the instability for $\zeta = 0$, and 2) we have
  been unable to find a simple test case that so strongly reproduces
  the instabilities. The continuum is optically thin in absorption
($\tau_{\rm abs} \approx 4\alog{-4}$) and reaches a scattering optical
depth $\tau_{\rm scatt} \approx 1.5$. We use a Planck-function as the
inner boundary condition for the intensities for simplicity,
physically it is better to employ a nebular [symmetry] boundary
condition which is shown in Figure~\ref{wiggles_cmf_neb}. We see that
the instability is evident (for the $\zeta=0$ case) even though the
line has gone into emission (note that the different scales
  between Fig.~\ref{wiggles_cmf} and Fig.~\ref{wiggles_cmf_neb} give
  the impression that the instability is somewhat suppressed in the
  nebular case, but it is simply due to the larger range required to
  plot the strong emission features). The unstable wiggles are not
  associated with the grid and varying the wavelength resolution does
  not significantly affect the output spectrum.  While we
  believe from our tests that the instability is produced
  predominantly in optically thin atmospheres, we have not been able
  to ascertain the precise conditions that trigger the instability.

The instability in the $\zeta=0$ discretization is obvious starting at
the rest wavelength of the line. Even in the line trough the
oscillations created by the instability are apparent. The overall line
shapes are distorted, even in the red emission feature. The
instability vanishes 
for $\zeta\ne 0$ as shown in the Fig.~\ref{wiggles_cmf}.  The
differences between the spectra for $\zeta=0.5$ and $1.0$ are very
small, the $\zeta=0.1$ case shows a few percent difference compared
to $\zeta=1$. It is clear that the instability of the $\zeta=0$ case
can easily be avoided by using $\zeta>0$ due to the stabilizing
features of the $\zeta=1$ discretization. In our test case even
$\zeta=0.1$ is sufficient to prevent the instability.

\section{Conclusions}

We have studied the discretization of the
$\frac{\partial}{\partial\lambda}$ term in the co-moving frame
radiative transfer equation. We have found that we can write a second
order discretization scheme, which is in general more accurate than
our previous method, but
it is possible for this scheme to become unstable. We have constructed a
hybrid Crank-Nicholson 
like scheme, which is unconditionally stable. We have been unable to
identify the exact conditions which lead to the unstable behavior, but
we have shown that even with a small admixture of the unconditionally
stable scheme, stability is recovered.  We recommend using a small
value of $\zeta = 0.1$, which in all our tests leads to recovered
stability and is more accurate than a higher value of
$\zeta$. However, until we determine the exact conditions that trigger
the instability the cautious user should vary $\zeta$ and determine
the sensitivity of the results to its value.

\begin{acknowledgements}
We thank the referee for a very helpful report which significantly
improved the presentation of this paper.
This work was supported in part by 
NASA grants NAG 5-8425 and NAG 5-3619 to
the University of Georgia and by NASA grant
NAG5-3505, NSF grants AST-0204771 and AST-0307323, and an IBM SUR
grant to the University of Oklahoma. PHH was 
supported in part by the P\^ole Scientifique de Mod\'elisation
Num\'erique at ENS-Lyon. Some of the calculations presented here were
performed at the San Diego Supercomputer Center (SDSC), supported by
the NSF, at the National Energy Research Supercomputer Center
(NERSC), supported by the U.S. DOE, and at the H\"ochstleistungs
Rechenzentrum Nord (HLRN).  We thank all these institutions 
for a generous allocation of computer time.
\end{acknowledgements}

\clearpage



\end{document}

%% file: macros.tex
%
%
\def\valid{}    

\font\caps=cmcsc10                  
\font\dunh=cmdunh10  at 12.0 true pt 
\font\dunhs=cmdunh10 
\font\vbold=cmbx10 scaled \magstep1 
\font\sevenbf=cmbx7
\font\sevenit=cmti7
\font\Kapi=cmr17

\def\MEV{DOME}
\def\RTE{equation of radiative transfer}
\def\etal{{et al}}
\def\HW{H\&W}
\def\OK{O\&K}
\def\ok{O\&K}
\def\RH{R\&H}

\def\ibmrs{\hbox{\tt RS/6000}}
\def\hp{\hbox{\tt HP~9000}}
\def\dec{\hbox{\tt DEC~5000}}
\def\axp{\hbox{\tt AXP}}
\def\ibmmf{\hbox{\tt IBM~3090}}
\def\ibmpc{\hbox{\tt 486DX}}
\def\cray{\hbox{\tt Cray 2}}
\def\ymp{\hbox{\tt YMP}}
\def\nec{\hbox{\tt NEC}}

\def\g{\gamma}
\def\b{\beta}
\def\m{\mu}
\def\e{\epsilon}
\def\n{\nu}
\def\l{\lambda}
\def\L{\Lambda}
\def\t{\tau}
\def\pder#1#2{{\partial #1 \over \partial #2}}
\def\div#1#2{{#1\over #2}}
\def\rout{\ifmmode{r_{\rm out}}\else\hbox{$r_{\rm out}$}\fi}
\def\tmax{\ifmmode{\tau_{\rm max}}\else\hbox{$\tau_{\rm max}$}\fi}
\def\tstd{\ifmmode{\tau_{\rm std}}\else\hbox{$\tau_{\rm std}$}\fi}
\def\vmax{\ifmmode{v_{\rm max}}\else\hbox{$v_{\rm max}$}\fi}
\def\muE{\ifmmode{\mu_{\rm E}}\else\hbox{$\mu_{\rm E}$}\fi} 
\def\pE{\ifmmode{p_{\rm E}}\else\hbox{$p_{\rm E}$}\fi} 
\def\bmax{\ifmmode{\b_{\rm max}}\else\hbox{$\b_{\rm max}$}\fi}
\def\kms{\hbox{$\,$km$\,$s$^{-1}$}}
\def\ergs{\hbox{$\,$erg$\,$s$^{-1}$}}
\def\kpc{\hbox{$\,$kpc} }
\def\ang{\hbox{\AA}}
\def\Msun{\hbox{$\,$M$_\odot$} }
\def\Lsun{\hbox{$\,$L$_\odot$} }
\def\Teff{\hbox{$\,T_{\rm eff}$} }
\def\alog#1{\times 10^{#1}}
\def\rin{\hbox{$r_{\rm in}$} }
\def\rout{\hbox{$r_{\rm out}$} }

\def\lstar{\ifmmode{\Lambda^*}\else\hbox{$\Lambda^*$}\fi} 
\def\Rop{\ifmmode{[R_{ij}]}\else\hbox{$[R_{ij}]$}\fi}
\def\Rij{\Rop}
\def\Rji{\ifmmode{[R_{ji}]}\else\hbox{$[R_{ji}]$}\fi}
\def\Rstar{\ifmmode{[R_{ij}^*]}\else\hbox{$[R_{ij}^*]$}\fi}
\def\Rijstar{\Rstar}
\def\Rjistar{\ifmmode{[R_{ji}^*]}\else\hbox{$[R_{ji}^*]$}\fi}
\def\DRji{\ifmmode{[\Delta R_{ji}]}\else\hbox{$[\Delta R_{ji}]$}\fi}
\def\DRij{\ifmmode{[\Delta R_{ij}]}\else\hbox{$[\Delta R_{ij}]$}\fi}

\def\Jb{{\bar J}}
\def\Jnew{{\bar J_{\rm new}}}
\def\Jold{{\bar J_{\rm old}}}
\def\Jfs{{\bar J_{\rm fs}}}
\def\Snew{{S_{\rm new}}}
\def\Sold{{S_{\rm old}}}
\def\Amat{\mat{A}}             

\def\ns{\ifmmode{N_{\rm s}}          
        \else\hbox{$N_{\rm s}$}\fi}
\def\ion#1{\hbox{ #1}}         

\def\peq{\mathbin{\hbox{$+$}\hbox{$=$}}}

\def\mat#1{{\bf #1}}     
\def\vek#1{{#1}}         

\newcount\eqcount
\eqcount=0
\def
  \nummer{
    \global\advance\eqcount by 1
    (\the\eqcount)
  }

\def
  \numadv{
    \global\advance\eqcount by 1
  }

\def
   \numout#1{
     (\the\eqcount #1)
  }

\def\ivek#1#2{\ifmmode{\vek{I}^{#1}_{#2}}
        \else\hbox{$\vek{I}^{#1}_{#2}$}\fi}

\def\ip#1{\ivek{+}{#1}}      
\def\im#1{\ivek{-}{#1}}      

\def\tmat#1#2{\ifmmode{\mat{t}^{#1}_{#2}}
        \else\hbox{$\mat{t}^{#1}_{#2}$}\fi}
\def\rmat#1#2{\ifmmode{\mat{r}^{#1}_{#2}}
        \else\hbox{$\mat{r}^{#1}_{#2}$}\fi}
\def\bvek#1#2{\ifmmode{\beta^{#1}_{#2}}
        \else\hbox{$\beta^{#1}_{#2}$}\fi}

\def\tpi#1{\tmat{+}{#1}}
\def\tmi#1{\tmat{-}{#1}}
\def\rmi#1{\rmat{-}{#1}}
\def\rpi#1{\rmat{+}{#1}}
\def\bpi#1{\bvek{+}{#1}}
\def\bmi#1{\bvek{-}{#1}}

\def\tp{\tmat{+}{}}          
\def\tm{\tmat{-}{}}          
\def\rmm{\rmat{-}{}}         
\def\rp{\rmat{+}{}}          
\def\bp{\bvek{+}{}}          
\def\bm{\bvek{-}{}}          
\def\tpm{\tmat{\pm}{}}       
\def\rpm{\rmat{\pm}{}}       
\def\bpm{\bvek{\pm}{}}       

\def\lp{\ifmmode{\lambda^+_\tau}           
        \else\hbox{$\lambda^+_\tau$}\fi}
\def\lm{\ifmmode\lambda^-_\tau             
        \else\hbox{$\lambda^-_\tau$}\fi}